\documentclass[a4papaer,11pt]{article}
\usepackage{graphicx}
\begin{document}

\title{$XMM$-$Newton$ Observations of G32.45+0.1 and G38.55+0.0:
diffuse hard X-ray sources found with the $ASCA$ Galactic Plane Survey
}

\author{Hiroya Yamaguchi$^1$,
	Masaru Ueno$^1$,
	Katsuji Koyama$^1$, \\
	Aya Bamba$^2$,
	and Shigeo Yamauchi$^3$ 
}

%
%




\maketitle

\begin{abstract}
We report on $XMM$-$Newton$ observations of  
G32.45+0.1 and G38.55+0.0.
These were discovered as diffuse hard X-ray sources
with the $ASCA$ Galactic plane survey, 
but the limited spatial resolution of $ASCA$
could not conclude whether these are truly diffuse or a group of unresolved 
point-sources.
$XMM$-$Newton$, with higher spatial resolution than $ASCA$,
confirmed that G32.45+0.1 has a diffuse shell-like structure 
with a radius of $\sim 4^{\prime}$.
The spectrum shows featureless continuum, hence
can be fitted with a power-law model of $\Gamma \sim 2.2$ 
with an absorption of  $N_{\rm H}\sim 5.2\times 10^{22}~{\rm cm}^{-2}$.
From this $N_{\rm H}$ value, we estimate the distance to G32.45+0.1 to be
$\sim 17$~kpc, then the luminosity (in the 0.5--10.0~keV band)  
and radius of the shell
are $\sim 9.5\times 10^{34}$~ergs~s$^{-1}$ and $\sim $20~pc, respectively.
The radio complex sources in the NRAO/VLA Sky Survey  (NVSS; 1.4~GHz)
are globally associated to the X-ray shell of G32.45+0.1.
Therefore G32.45+0.1 is likely to be a synchrotron dominant
shell-like SNR.
No significant diffuse structure was found
in  the $XMM$-$Newton$ image of another $ASCA$ diffuse source G38.55+0.0. 
The upper limit of the observed flux (0.5--10.0~keV) is  estimated to be
$9.0\times 10^{-13}$~ergs~cm$^{-2}$~s$^{-1}$ 
being consistent with the $ASCA$ result.

\footnote[1]{Department of Physics, Graduate School of Science,
	Kyoto University, Sakyo-ku, Kyoto 606-8502}
\footnote[2]{RIKEN (The Institute of Physical and Chemical Research),
	2-1 Hirosawa, Wako, Saitama 351-0198}
\footnote[3]{Faculty of Humanities and Social Sciences,
	Iwate University, 3-18-34 Ueda, Morioka, Iwate 020-8550}

\end{abstract}

\section{Introduction}

Since cosmic rays were discovered by Hess in 1912,
the source and the mechanism of acceleration have been unknown.
The spectrum of cosmic rays shows a single power-law 
up to the knee energy ($\sim 10^{15.5}$~eV),
where the gyro radius of electrons 
in typical interstellar magnetic field
is much smaller than the Galactic radius. 
Therefore  cosmic rays below the
knee energy is likely  to be Galactic origin.
The discoveries of synchrotron X-rays and 
inverse Compton TeV $\gamma $-rays from the supernova remnant (SNR) 
SN~1006 indicated that electrons are accelerated 
close to the knee energy (Koyama et al. 1995; Tanimori et al. 1998).
The $Chandra$ results of the small-scale structure 
is successfully explained 
by a diffusive shock acceleration model (DSA); 
the first-order Fermi mechanism is working at the shock front of SNRs
(Bamba el al. 2003b; Yamazaki et al. 2004).

Besides SN~1006, synchrotron X-ray emissions were discovered 
from other SNRs, 
G347.5-0.3 (RX~J1713.7$-$3946; Koyama et al. 1997; Slane et al. 1997),
Cas~A (Vink et al. 2000), 
Tycho's SNR (Hwang et al. 2002), 
RX~J0852.0$-$4622 (Slane et al. 2001), 
RCW~86 (Borkowski et al. 2001a),  
AX~J1843.8$-$0352 (Bamba et al. 2001; Ueno et al. 2003),
30~Dor~C (Bamba et al. 2004).
Although the synchrotron X-ray emissions were observed from several SNRs,
the total number and the X-ray fluxes so far discovered are insufficient
to account for all cosmic rays in our Galaxy; if SNRs are 
the main accelerator of the cosmic 
rays, we can expect more non-thermal SNRs. 
Accordingly, we searched for the SNR candidates in the 
data of the $ASCA$ Galactic plane survey which covered  
$|l|\leq 45^{\circ}$ 
$|b|\leq 0.^{\circ}4$ on the Galactic plane (Yamauchi et al. 2002).
About half a dozen candidates were found in this survey area;
follow-up deep exposure observations with $ASCA$ were made 
on three candidates, G11.0+0.0, G25.5+0.0, 
and G26.6$-$0.0, and these were suggested  to be  non-thermal SNRs 
(Bamba et al. 2003a).
G28.6$-$0.1 (AX~J1843.8$-$0352) is deeply observed with 
both $ASCA$ and $Chandra$ and is established to be
a synchrotron X-ray emitting shell-like SNR 
(Bamba et al. 2001; Ueno et al. 2003).

Two other candidates G32.45+0.1 and G38.55+0.0 are 
seen as diffuse-like hard sources  in the $ASCA$ images, 
and the spectra show a power-law like feature with $\Gamma $=1--3.
The spatial sizes ($\sim$ a few arcmin) are, however, 
marginal compared with the $ASCA$
point spread function (PSF) of ($\sim$ 1~arcmin) 
to determine whether these are diffuse or 
a group of unresolved point-sources.
We, therefore, made $XMM$-$Newton$ (Aschenbach et al. 2000) observations
on  G32.45+0.1 and G38.55+0.0 with higher spatial resolution 
(PSF$\sim $5~arcsec) 
and larger effective area than those of $ASCA$.

\section{Observations}

G32.45+0.1 and G38.55+0.0 were
observed with $XMM$-$Newton$ on 2003 September 25 
(Observation ID~=~0136030101), 
and 2003 September 21 (Observation ID~=~0136030201), respectively.
Although the data were obtained from both 
the EPIC-MOS (Turner et al. 2001)
and EPIC-PN (Str$\ddot{\rm u}$der et al. 2001) cameras, 
the major part of these  objects were suffered 
by the PN-CCD gaps and bad columns, 
and we verified that cleaned PN data do not improve 
accuracy of results obtained with MOS only,
hence  we ignore the PN data hereafter.
All the EPIC instruments were operated in the full frame mode 
with the medium filter.
We used version 5.4.1 of the Standard Analysis System (SAS)
software, and selected X-ray events with PATTERN keywords between 0 and 12.

The net exposure times were 26.8~ks and 15.9~ks for 
G32.45+0.1 and G38.55+0.0, respectively.
In the observation of G32.45+0.1, however, 
the particle background was exceptionally high and variable, 
and hence we accumulated the background light-curve in the 10--12~keV band
from all of the field of view and removed  the  high background data  
(the time intervals when the count rate is larger than 0.3~count~s$^{-1}$).
After the filtering, the exposure times of MOS1 and MOS2 are
20.5~ks and 21.2~ks, respectively.
The filtering using the same prescription was also performed 
on the observation of G38.55+0.0.
Then the filtered exposure times of MOS1 and MOS2 are 15.1~ks 
and 15.4~ks, respectively.

In order to estimate the residual soft-proton cotamination 
(which is vignetted),
we calculated the value ``R'' according to Section B.2 of 
De Luca and Molendi (2004).
For both of the observations, the values R of MOS1 and MOS2 are 
1.2 and 1.1, respectively, 
hence the contributions of residual
soft-protons are negligible 
in the filtered data.

\section{Analysis of G32.45+0.1}

\subsection{X-Ray Image}

\begin{figure}[h]
  \begin{center}
    \includegraphics[width=9cm,keepaspectratio=true,clip]{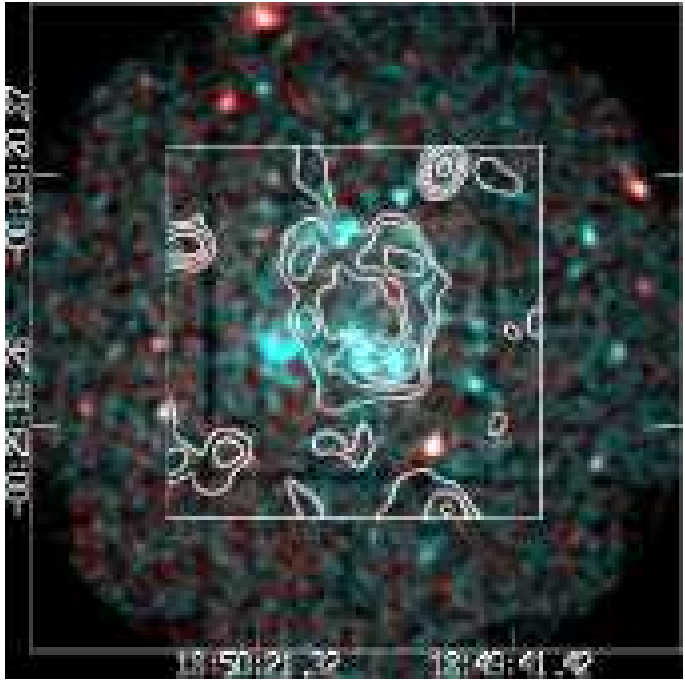}
  \end{center}
  \caption{
	$XMM$-$Newton$/MOS1+2 image of G32.45+0.1 
	overlaid on the NVSS 20~cm radio contours 
	which are expressed on the inner region of the square
	(Condon et al. 1998).
	The soft (0.5--2.0~keV) and hard (2.0--7.0~keV) band images 
	are represented with red and blue colors, 
	where the background events are not subtracted.
	The image has been smoothed to a resolution of $9.^{\prime \prime}6$.
	The  coordinate (R.A. Dec) are  J2000.}

  \label{fig:G32_2color_img}
\end{figure}

In figure \ref{fig:G32_2color_img}, 
the soft (0.5--2.0~keV) and hard (2.0--7.0~keV) band 
X-ray images are shown with two colors.
In the hard band image (blue) only, we see a  shell-like structure 
with a radius of about 4~arcmin.

\begin{figure}[htbp]
  \begin{center}
    \includegraphics[width=9cm,keepaspectratio=true,clip]{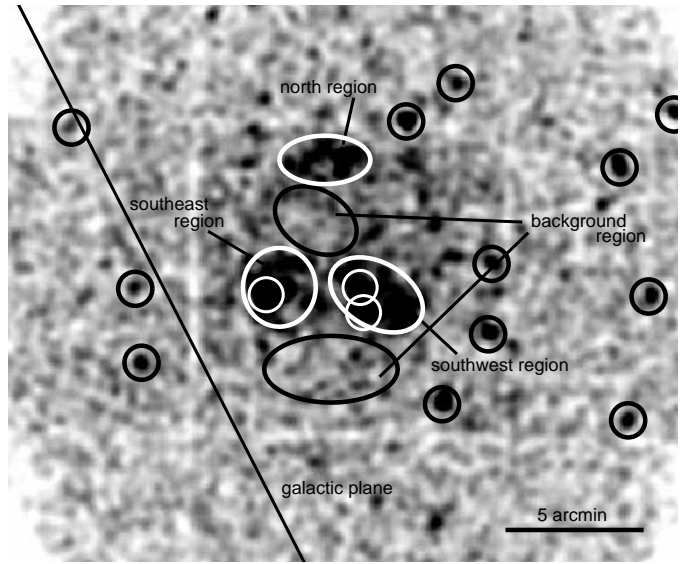}
  \end{center}
  \caption{
	The $XMM$-$Newton$/MOS1+2 hard band image 
	in the 2--7~keV band.
	The background events are not subtracted.
	The solid line means the Galactic plane.
	Three (bold) white ellipses and two black ellipses 
        are source and  background regions, respectively.
	Point-sources are expressed as the black small circles.}

  \label{fig:region}
\end{figure}

In order to estimate the accurate flux
of the diffuse shell in the hard band image, 
we picked up point-sources  using the EWAVELET software,
with the detection threshold of 5$\sigma $.
Thus detected point-sources are indicated  
with the black and (thin) white circles in figure \ref{fig:region}.
The three white circles are in the diffuse shell, and hence
we made radial profiles of these sources and estimated
the FWHM assuming Gaussian profile. 
As listed  in table \ref{tab:extent},
all the FWHMs of the sources are significantly larger than 
that of the point spread function (PSF)  of $XMM$-$Newton$ 
(FWHM $\sim 5^{\prime \prime}$).
We, therefore, conclude that these are not point-like sources, but 
are local enhancements on the shell.

\begin{table}[hp]
 \begin{center}
  \caption{The position and the extension of 
	the point-like sources  in the shell 
	(white circles in figure \ref{fig:region}).
}
  \label{tab:extent}
  \begin{tabular}{ccc}
	\hline \hline
	(R.A., Dec.)$_{\rm J2000}$ & position & extension (FWHM arcsec) \\
	\hline
	($18^{\rm h}50^{\rm m}19^{\rm s}$,
		~$-00^{\circ}26^{\prime}20^{\prime \prime}$) 
		& the southeast region & $60\pm 25$ \\
	($18^{\rm h}50^{\rm m}06^{\rm s}$,
		~$-00^{\circ}25^{\prime}58^{\prime \prime}$) 
		& the southeast region & $60\pm 25$ \\
	($18^{\rm h}50^{\rm m}05^{\rm s}$,
		~$-00^{\circ}26^{\prime}58^{\prime \prime}$) 
		& the southeast region & $26\pm 6.1$ \\
	\hline
	\multicolumn{3}{l}{Uncertainties correspond to 1$\sigma $ regions.}
  \end{tabular}
 \end{center}
\end{table}

\subsection{Spectrum}

\begin{figure}[hp]
  \begin{center}
	\rotatebox{270}{\includegraphics[width=6cm,keepaspectratio=true,clip]{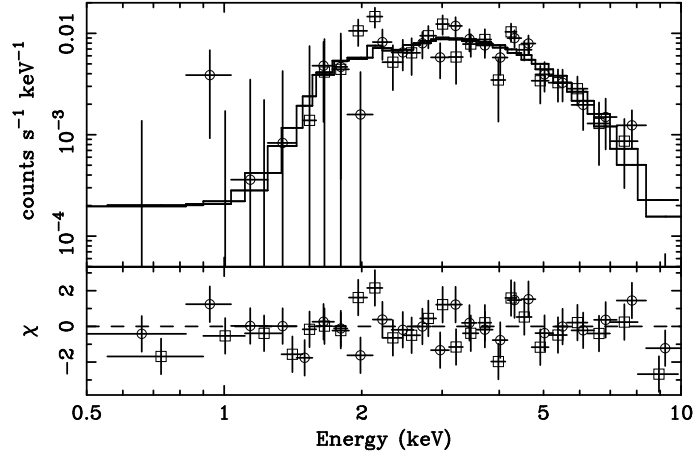}}
  \end{center}
  \caption{
	X-ray spectra of G32.45+0.1 observed 
	with MOS1 (circle) and MOS2 (square).
	The solid lines show the best-fit power-low models.
	}
  \label{fig:spec}
\end{figure}

Firstly, we divided the diffuse emission into three regions
(north, southwest, and southeast; figure \ref{fig:region}), 
and extracted the X-ray spectra separately.
However the statistics of all the spectra are 
not enough for spectral fittings.
We therefore added them together and 
regarded it as the spectrum of the whole shell.
The background are taken from the neighborhood ellipses
which are also shown in figure \ref{fig:region}.
Since these background regions are near to 
the source regions enough 
and the average off-axis angle of them is 
almost equal to that of source regions,
the contribution of the vignetting effect is negligible 
in this spectral measurements,
and hence we did not apply the vignetting correction.
The background-subtracted spectra of MOS1 and MOS2
are shown in figure \ref{fig:spec}.
Since the spectra have no emission line,
we fitted them with a model of power-law modified 
by photo-electric absorption .
The cross-sections were taken from 
Morrison and McCammon (1983).
The fittings were statistically accepted for the spectra
with the best-fit parameters shown in table \ref{tab:G32_param}.

For comparison, we tried fittings 
with a thin thermal plasma model
in non-equilibrium ionization
(an NEI model; Borkowski et al. 2001b),
and obtained a nearly equal fit.
The best-fit parameters of the NEI model are shown in table 2.

 \begin{table}[h]
 \begin{center}
  \caption{Results of spectral fitting of G32.45+0.1$^a$}
  \label{tab:G32_param}
  \begin{tabular}{lcc}
	\hline \hline
	Parameters & Power-law & NEI  \\  
	\hline
	$N_{\rm H} ~[\times 10^{22}~{\rm cm}^{-2}]$ 
		& 5.2 (3.9--7.5) & 4.6 (3.4--6.1) \\
	$\Gamma / kT$~[keV]  
		& 2.2 (1.8--3.0) & 5.2 (3.0--12) \\
	Abundance$^b$ 
		& $--$ 		 & $1.0\times 10^{-2}$ ($<$ 0.35) \\
	log($n_et$)~[cm$^{-3}$~s]
		& $--$ 		 & $3.7\times 10^{13}~^c$  \\
	Flux$^d ~[{\rm ergs}~{\rm cm}^{-2}~{\rm s}^{-1}]$ 
		& $2.8\times 10^{-12}$   &   $1.9\times 10^{-12}$   \\
	Reduced $\chi ^2$ (d.o.f.)  
		& 1.14 (46) 	 & 1.19 (44) \\
	\hline
   \multicolumn{3}{l}{$^a$Parentheses indicate 
			single parameter 90\% confidence intervals.} \\
   \multicolumn{3}{l}{$^b$Assuming the solar abundance ratio 
		(Anders, Grevese 1989).} \\
   \multicolumn{3}{l}{$^c$The error could not be decided 
			in the region of (1$\times 10^8$--5$\times 10^{13}$)} \\ 
   \multicolumn{3}{l}{\  which is the limit of XSPEC software.} \\
   \multicolumn{3}{l}{$^d$Absorption corrected flux in the 0.5--10.0~keV band.}
  \end{tabular}
 \end{center}
\end{table}

\section{Analysis of G38.55+0.0}

\begin{figure}
  \begin{center}
    \includegraphics[width=6cm,keepaspectratio=true,clip]{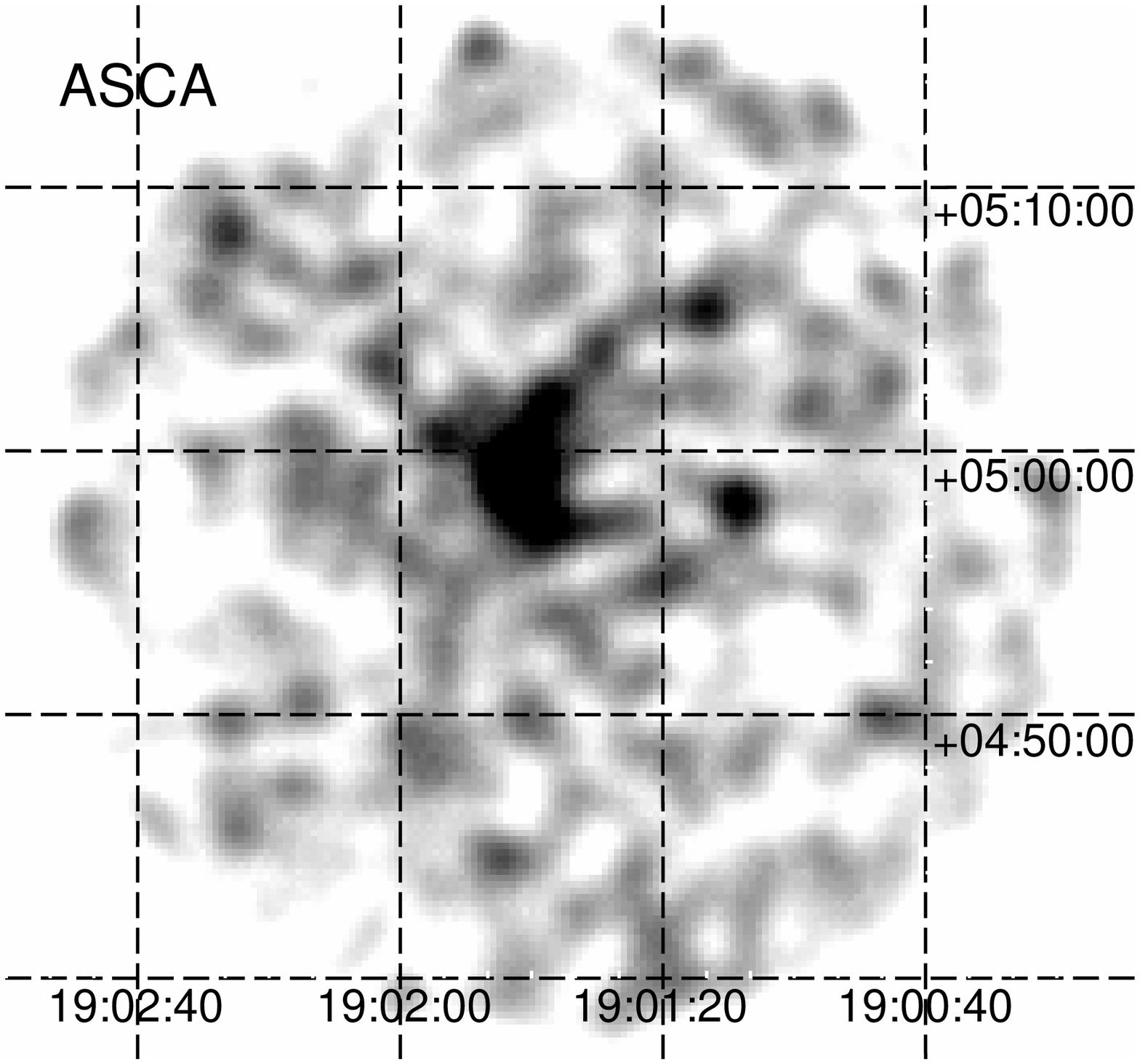}
    \includegraphics[width=6cm,keepaspectratio=true,clip]{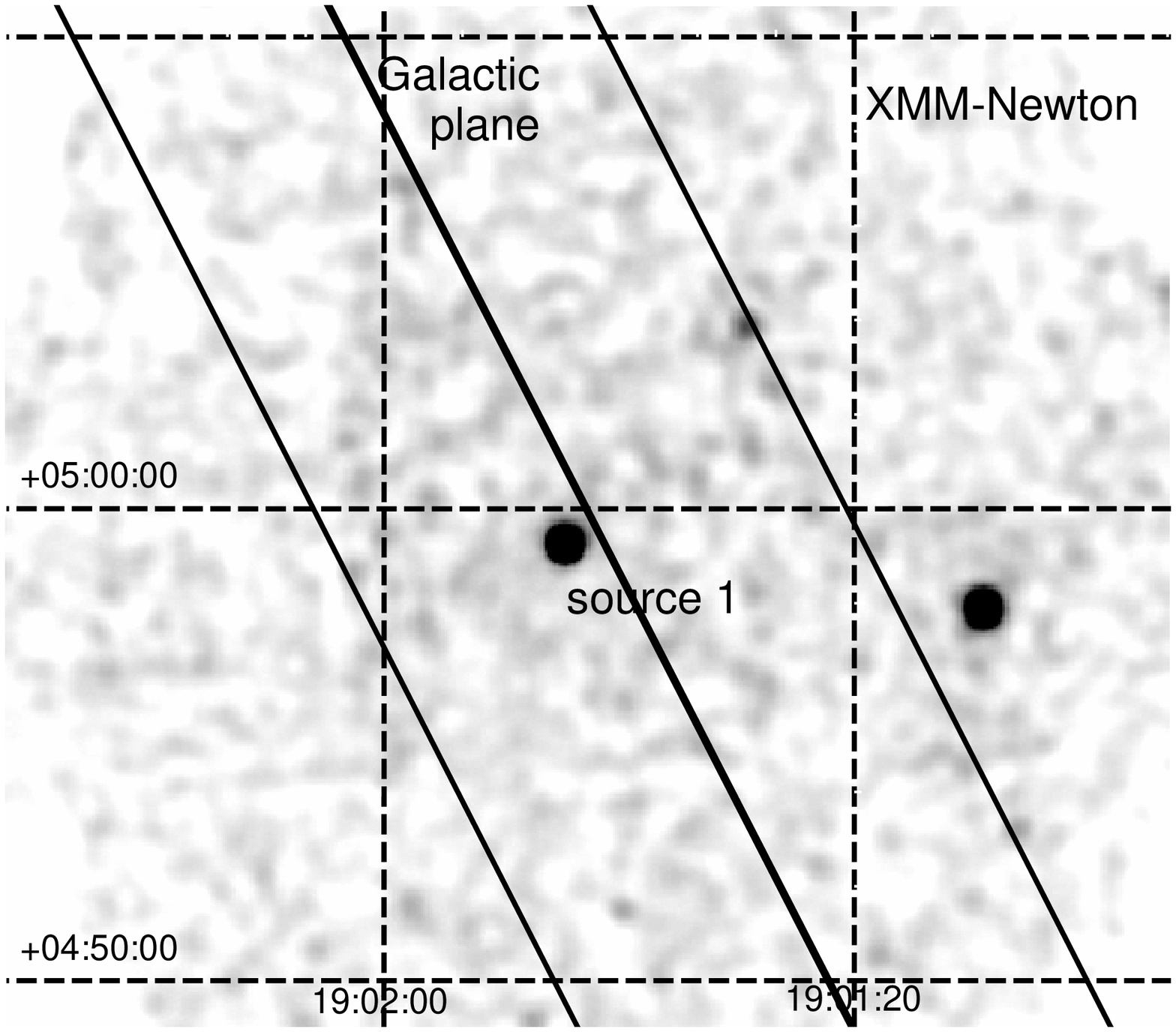}
  \end{center}
  \caption{$ASCA$/GIS2+3 and $XMM$-$Newton$/MOS1+2 images of 
	the G38.55+0.0 region in the hard X-ray band 
	where the background events are not subtracted.
	The axis labels represent the R.A. Dec. (J2000) coordinates.
	The center line and the two outer lines 
	expressed in the right panel
	mean the Galactic plane and
	$b=\pm 5^{\prime}$, respectively.	
}
  \label{fig:G38img}
\end{figure}

The EPIC/MOS1+2 image in the 2.0--7.0~keV band near
G38.55+0.0 is shown  in the right panel of figure \ref{fig:G38img}.
For comparison, we show the $ASCA$ image in the same energy band 
(2.0--7.0~keV band)
near the  G38.55+0.0 region in the  left panel of figure \ref{fig:G38img}.
Although, we can see a structure extending larger than  the
point spread function (PSF) of $ASCA$ ($\sim 1^{\prime}$),  no 
significant diffuse emission except some point-sources are found
in the $XMM$-$Newton$ image.  A possibility of a group of
unresolved  point-sources may be excluded, because only one
point-source is found near the center of G38.55+0.0 by
the $XMM$-$Newton$ observation  of PSF$\sim 5^{\prime \prime}$ (FWHM).
We first extracted the spectrum of the  point-source (source~1)
at the center of G38.55+ 0.0. The source region was selected from a 
$25^{\prime \prime}$-radius circle, and the background was taken  
from an annular region  around the source with the inner 
and outer radii of $25^{\prime \prime}$ and $100^{\prime \prime}$.  
We fitted the background-subtracted spectrum with a power-law model.
The fitting was statistically accepted with 
the best-fit parameters shown in table \ref{tab:G38_param}.
For comparison, we made the $ASCA$ spectrum from a $6^{\prime}$-radius region
with a background from an annulus of  inner-outer radius of
$6^{\prime}$--$12^{\prime}$.  
The background subtracted spectrum is also fitted with a power-law model.
The best-fit parameters are given in table 3.
Within the statistical error, we see no significant 
difference of the best-fit $N_{\rm H}$ and $\Gamma $ for
the $ASCA$ ``diffuse" emission and the $XMM$-$Newton$ point-source emission.
However the observed flux of $ASCA$ is $\sim $5 times larger than  
that of source~1
in the $XMM$-$Newton$ image.  
Therefore one possibility is that there is a ``diffuse'' component 
with low surface brightness.

\begin{table}
 \begin{center}
  \caption{Results of spectral fittings of source~1
	and the $ASCA$ ``diffuse'' emission.$^a$}
  \label{tab:G38_param}
  \begin{tabular}{lcc}
	\hline \hline
	Parameter & source~1 ($XMM$-$Newton$) 
		& $ASCA$ 6'-radius region \\
	\hline
	$N_{\rm H} [\times 10^{22}~{\rm cm}^{-2}]$ & 
	3.3 (2.0--5.3) & 2.0 (1.6--2.6)  \\
	$\Gamma $ & 1.7 (1.1--2.7) & 1.1 (0.9--1.3) \\
	Observed flux$^b [{\rm ergs}~{\rm cm}^{-2}~{\rm s}^{-1}]$
	 & $2.6 (2.4-2.9)\times 10^{-13}$ & $1.2 (0.82-1.6)\times 10^{-12}$ \\
	$N_{\rm H}$ corrected flux$^b [{\rm ergs}~{\rm cm}^{-2}~{\rm s}^{-1}]$ 
	 & $5.1 (4.6-5.7)\times 10^{-13}$ & $1.6 (1.1-2.1)\times 10^{-12}$ \\
	Reduced $\chi ^2 $ (d.o.f.)  & 0.295 (5) & 0.912 (19) \\
	\hline
   \multicolumn{3}{l}{$^a$Parentheses indicate 90\% confidence intervals.} \\
   \multicolumn{3}{l}{$^b$Flux in the 0.5--10.0~keV band.}
  \end{tabular}
 \end{center}
\end{table}

\begin{figure}[h]
  \begin{center}
	\includegraphics[width=6cm,keepaspectratio=true,clip]{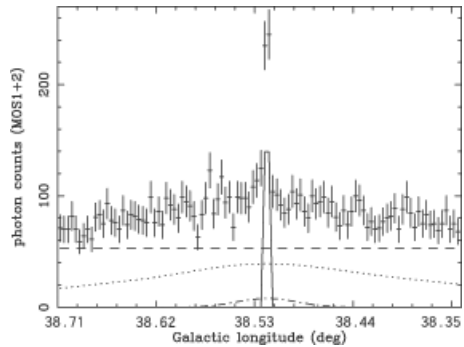}
  \end{center}
  \caption{
	The projected profile of the sum of MOS1 and MOS2
	photon counts in the 2--7~keV band 
	in the $|b|\leq 5^{\prime}$ 
        region along $l$ (Galactic longitude).
	Solid, dash, dotted, and dash-dotted lines show 
	the best-fit curves of point-source~1, NXB, CXB + GRXE, 
	and the putative ``diffuse'' emission, 
	respectively. 
	}
  \label{fig:profile}
\end{figure}

In order to estimate the flux of the putative ``diffuse'' emission 
in the $XMM$-$Newton$ observation,
we made a projected profile along the 
Galactic plane with the width of $|b|\leq 5^{\prime}$ 
using the hard band image (2--7~keV, see figure \ref{fig:G38img}). 
Then the profile is fitted with a model function made
under the following assumptions; 
the non-X-ray background (NXB) is 
constant in the $XMM$-field of view, the cosmic X-ray background (CXB)
and  the Galactic Ridge emission (GRXE) are also constant along the Galactic 
plane but are modified by the vignetting effect, 
and the contribution of source~1
is given by a Gaussian profile with the width of PSF. 
In addition, we approximated the putative ``diffuse'' emission
to be a simple Gaussian of 3$\sigma =5^{\prime}$ with the center at
the source~1 position (see the $ASCA$ image).
Figure \ref{fig:profile} is the projected profile and the best-fit model; 
NXB is shown with the dash line, the sum of CXB and GRXE 
is given with the dotted line,
and source~1 is shown with the solid line.  
With this fitting, the flux of the putative
``diffuse'' emission is constrained to be 
less than 0.018~counts~s$^{-1}$ (90\% upper limit),
so this component is undetected in the $XMM$-$Newton$ observation,
unlike $ASCA$ observation.
A reason for this may be higher particle background than 
that of $ASCA$ observation.

This value is converted to 
$ 6.4\times 10^{-13}$~ergs~cm$^{-2}$~s$^{-1}$ (in 0.5--10~keV),
under the power-law model with $\Gamma =1.1$ and 
$N_{\rm H}=2.0\times 10^{22}~{\rm cm}^{-2}$, 
the best-fit $ASCA$ parameters of G38.55+0.0.

\begin{table}[h]
 \begin{center}
  \caption{Results of projected profile fitting
	and flux estimation of the G38.55+0.0 region.
}
  \label{tab:upper_limit}
  \begin{tabular}{lcl}
	\hline \hline
	component & count rate$^a$ & model\\
	\hline
	NXB [counts~s$^{-1}$~deg$^{-2}$] 
		& 2.8~(1.8--4.0) & constant \\
	CXB + GRXE [counts~s$^{-1}$~deg$^{-2}$] & 
		2.3~(0.43--4.3) & constant$^b$ \\
	point-source [counts~s$^{-1}$] & 1.0~(0.68--1.3)$\times 10^{-2}$
					 & Gaussian \\
	diffuse [counts~s$^{-1}$] & $<0.018 $  & Gaussian \\
	\hline
  \multicolumn{3}{l}{$^a$All count rates are the photon rates in the 2--7~keV band.} \\
  \multicolumn{3}{l}{$^b$Constant along the Galactic plane,
			but is modified by the vignetting effect.} \\
  \multicolumn{3}{l}{Parentheses indicate 90\% errors.} 
  \end{tabular}
 \end{center}
\end{table}

\section{Discussion}

\subsection{G32.45+0.1}

We found a clear shell structure with a radius of $\sim 4^{\prime}$
in the $XMM$-$Newton$ hard band (2.0--7.0~keV) image.
The X-ray spectrum is well-fitted with either 
a power-law or a thin thermal plasma model.
The thermal model requires the temperature of 3.0--12~keV
which is similar to (or higher than) that of typical 
young SNRs such as Cas~A and Tycho's SNR 
($k$T $\sim $2--4~keV; e.g., Vink et al. 1996,
Gotthelf et al. 2001 (Cas~A); Fink et al. 1994 (Tycho's SNR)~).
This model, however, also requires an uncomfortably 
low metal abundance ($<$ 0.35~solar abundance),
compared with Cas~A and Tycho's SNR ($\sim $2~solar abundance).
Thus the spectrum of G32.45+0.1 is likely to be a power-law.
The best-fit parameter of 
$N_{\rm H}$=5.2~(3.9--7.5)$\times 10^{22}~{\rm cm}^{-2}$ 
gives rough estimation of the source distance to be 17~kpc 
with an assumption 
that the density in the Galactic plane is 1~H~cm$^{-3}$.
Then the X-ray luminosity (0.5--10.0~keV) and the radius are estimated
to be $\sim 9.5\times 10^{34}$~ergs~s$^{-1}$ and $\sim 20$~pc, respectively.
These values are not largely different from 
those of the non-thermal component of SN~1006 
( $\sim 2.0\times 10^{34}$~ergs~s$^{-1}$
and $\sim $9.5~pc
; e.g., Dyer et al. 2004).

We found a radio shell at the position of G32.45+0.1
from the NRAO/VLA Sky Survey (NVSS) data at 1.4~GHz (Condon et al. 1998) 
overplotted with contours in figure \ref{fig:G32_2color_img},
although the authors regarded  the radio shell as 
7 individual (point-like) sources 
as listed in the NVSS catalogue. 
The  flux sum is $\sim 1.7\times 10^{-1}$~Jy.
From the shell-like morphology in both the radio and X-ray bands
and the X-ray power-law model of photon index of $\Gamma $=2.2~(1.8--3.0),
G32.45+0.1 is highly possible to be a synchrotron X-ray emitting 
shell-type SNR.
For a discussion of a wide band spectrum,
we fitted the X-ray spectrum with an SRCUT model 
(Reynolds and Keohane 1999).
The radio index has not yet been determined, 
but simple acceleration model predicts
the energy index to be 0.5.
In fact, the energy index ($\alpha $) at 1~GHz 
of SN~1006, the typical synchrotron X-ray SNR, 
is reported to be $\sim $0.57 (Allen et al. 2001).
Thus we tried fitting assuming $\alpha $ to be 0.5 or 0.6.
The best-fit parameters are shown in table \ref{tab:srcut}.

\begin{table}[h]
 \begin{center}
  \caption{The best-fit parameters by the SRCUT model$^a$}
  \label{tab:srcut}
  \begin{tabular}{lcc}
	\hline \hline
	Parameters & $\alpha = 0.5^b$ & $\alpha = 0.6^b$ \\  
	\hline
	$N_{\rm H} [\times 10^{22}~{\rm cm}^{-2}]$ 
		& 5.2 (4.5--6.2) & 5.2 (4.4--6.1) \\
	Cut-off frequency [$\times 10^{17}$~Hz]  
		& 3.4 (0.79--91) & 5.4 ($<$ 35) \\
	Norm$^c$ & 0.019 (0.017--0.022) & 0.11 (0.095--0.12)  \\
	Reduced $\chi ^2$ (d.o.f.)  & 1.14 (46) & 1.14 (46) \\
	\hline
   \multicolumn{3}{l}{$^a$Parentheses indicate 90\% confidence intervals.} \\
   \multicolumn{3}{l}{$^b$The fixed value of $\alpha $.} \\
   \multicolumn{3}{l}{$^c$The flux density at 1~GHz~[Jy].}
  \end{tabular}
 \end{center}
\end{table}

Here the absorption is consistent with the fitting result with 
the power-law model.
Using the values of the cut-off frequency, 
we can estimate the maximum energy of electrons 
accelerated by the SNR shock 
according to Reynolds and Keohane (1999).
Assuming the cut-off frequency is $5\times 10^{17}$~Hz,
the maximum electron energy ($E_{\rm max}$) is estimated to be 
\[
E_{\rm max} \sim 60
\left( \frac{10\mu {\rm G}}{B} \right) ^{\frac{1}{2}} {\rm TeV.}
\]
This result is consistent with the standard picture of 
the non-thermal SNRs. 
For example, $E_{\rm max}$ and $B$ of SN~1006 are 20--70~TeV and 
3.5--85~$\mu $G, respectively (Bamba et al. 2003b).

In either $\alpha =$ 0.5 or 0.6,
the predicted flux density at 1.4~GHz is smaller than NVSS result.
This apparent inconsistency may be attributable
to a non-uniform shell structure.
In fact,  the radio brightest region  show no X-ray shell 
(see figure \ref{fig:G32_2color_img}).
This can be  interpreted that the X-ray emission is dominated
in relatively  low $B$ region, because the synchrotron energy loss rate is 
proportional to $B^2$, and hence the energy loss 
at the X-ray emitting electron energy
may be negligible. 
The radio shell, on the other hand, may  come from a higher $B$ region,
where energy loss of high energy electron is significant, 
and hence X-ray is weak.
More quantitative study requires spatial resolved X-ray and radio spectroscopy,
which is beyond the scope of this paper.

\subsection{G38.55+0.0}

On the contrary to the $ASCA$ results, no significant diffuse emission  is 
detected in the $XMM$-$Newton$  image of the G38.55+0.0 region.  
The absorbed surface brightness~(2.0--10.0~keV) of the GRXE is 
$\sim 3.5\times 10^{-11}~{\rm ergs}~{\rm cm}^{-2}~{\rm s}^{-1}~{\rm deg}^{-2}$ 
at $l=38.^{\circ}55$ (Sugizaki et al. 2001).
Assuming a thin thermal plasma of 7~keV for GRXE (Kaneda et al. 1997),
the $XMM$-$Newton$ count  is estimated to be 
$\sim 1.0$~counts~s$^{-1}$~deg$^{-2}$.
The unabsorbed surface brightness of the CXB component (in 2.0--10.0~keV), 
on the other hand, is 
$\sim 1.8\times 10^{-11}~{\rm ergs}~{\rm cm}^{-2}~{\rm s}^{-1}~{\rm deg}^{-2}$ 
(Ishisaki 1996). With the Galactic absorption of  
$N_{\rm H}=3.0\times 10^{22}$~cm$^{-2}$, the count rate is estimated to be 
$\sim 1.0$~counts~s$^{-1}$~deg$^{-2}$. Then the sum (GRXE+CXB) 
is $\sim 2.0$~counts~s$^{-1}$~deg$^{-2}$. 
This value is consistent with  the best-fit result of the projected profile 
in table \ref{tab:upper_limit} (2.3~counts~s$^{-1}$~deg$^{-2}$). 
Thus our flux estimation based on the model fitting to the projected profile 
is reliable. 
We then obtain the absorbed flux upper limit of 
the putative ``diffuse'' emission 
to be $6.4\times 10^{-13}$~ergs~cm$^{-2}$~s$^{-1}$ (0.5--10 keV band),
and the absorbed flux of 
the resolved point-source to be $2.6\times 10^{-13}$~ergs~cm$^{-2}$~s$^{-1}$
in the same energy band.
Then the sum of these fluxes is  $9.0\times 10^{-13}$~ergs~cm$^{-2}$~s$^{-1}$,
which is consistent with the $ASCA$ observed flux of G38.55+0.0. 
We note that no radio counterpart at the position of G38.55+0.0 has been
reported.  Accordingly, whether G38.55+0.0 is diffuse source or a new SNR 
is still an open issue.

\section{Summary}

The results of $XMM$-$Newton$ observations and analyses 
of G32.45+0.1 and G38.55+0.0 are summarized as follows:

\begin{enumerate}

\item G32.45+0.1 shows a clear shell-like structure 
	in the hard X-ray band.
	
\item The spectrum of G32.45+0.1 shows a non-thermal feature, and
	can be fitted with a power-law model of 
	$\Gamma \sim $2.2, 
	which suggests synchrotron X-ray emission from the shell of the SNR.
	
\item   The $N_{\rm H} \sim 5.2\times 10^{22}~{\rm cm}^{-2}$ gives the source 
	distance to be 17~kpc.
        Then the X-ray luminosity in the 0.5--10.0~keV band  
	and the shell radius 
	of G32.45+0.1 are  estimated to be 
	$\sim 9.5\times 10^{34}~{\rm ergs}~{\rm s}^{-1}$, and 20~pc, 
	respectively.

\item   No significant  diffuse emission from  G38.55+0.0 is detected.
	The upper limit in the 0.5--10.0~keV band is 
	$\sim 9.0\times 10^{-13}~{\rm ergs}~{\rm cm}^{-2}~{\rm s}^{-1}$, 
	consistent with the $ASCA$ flux.

\end{enumerate}

\bigskip

We thank all the members of the $ASCA$ Galactic plane survey team.
M.U. is supported by JSPS Research Fellowship for Young Scientists.
This work is supported by a Grant-in-Aid for the 21 century COE,
``Center for Diversity and Universality in Physics''.


\begin{thebibliography}{}


\bibitem -Allen, 
G.~E., Petre, R., \& Gotthelf, E.~V.\ 2001, ApJ, 558, 739 

\item Anders, E., \& Grevesse, N.\ 1989, Geochim. Cosmochim. Acta, 53, 197 

\bibitem -Aschenbach, B., 
Briel, U.~G., Haberl, F., Braeuninger, H.~W., Burkert, W., Oppitz, A., 
Gondoin, P., \& Lumb, D.~H.\ 2000, Proc. SPIE, 4012, 731 

\bibitem
-Bamba, A., Ueno, M., Koyama, K., \& Yamauchi, S.\ 2001, PASJ, 53, L21 

\bibitem
-Bamba, A., Ueno, M., Koyama, K., \& Yamauchi, S.\ 2003a, ApJ, 589, 253 

\bibitem
-Bamba, A., Yamazaki, R., Ueno, M., \& Koyama, K.\ 2003b, ApJ, 589, 827 

\bibitem 
-Bamba, A., Ueno, M., Nakajima, H., \& Koyama, K.\ 2004, ApJ, 602, 257 

\bibitem   
-Borkowski, K.~J., Rho, J., Reynolds, S.~P., \& Dyer, K.~K.\ 2001a, ApJ, 
550, 334 

\bibitem 
-Borkowski, K.~J., Lyerly, W.~J., \& Reynolds, S.~P.\ 2001b, ApJ, 
548, 820 

\bibitem -Condon, J.~J., Cotton, 
W.~D., Greisen, E.~W., Yin, Q.~F., Perley, R.~A., Taylor, G.~B., \& 
Broderick, J.~J.\ 1998, AJ, 115, 1693 

\bibitem -De Luca, A.~\& 
Molendi, S.\ 2004, A\&A, 419, 837 

\bibitem -Dyer, 
K.~K., Reynolds, S.~P., \& Borkowski, K.~J.\ 2004, ApJ, 600, 752 

\bibitem -Fink, H.~H., Asaoka, I., 
Brinkmann, W., Kawai, N., \& Koyama, K.\ 1994, A\&A, 283, 635 

\bibitem -Gotthelf, E.~V., 
Koralesky, B., Rudnick, L., Jones, T.~W., Hwang, U., \& Petre, R.\ 2001, 
ApJL, 552, L39 

\bibitem 
-Hwang, U., Decourchelle, A., Holt, S.~S., \& Petre, R.\ 2002, ApJ, 581, 
1101 

\item Ishisaki, Y. 1996, Ph.D. thesis, Univ. Tokyo

\bibitem -Kaneda, H., Makishima, 
K., Yamauchi, S., Koyama, K., Matsuzaki, K., \& Yamasaki, N.~Y.\ 1997, 
ApJ, 491, 638 

\bibitem -Koyama, K., Petre, R., 
Gotthelf, E.~V., Hwang, U., Matsuura, M., Ozaki, M., \& Holt, S.~S.\ 1995, 
Nature, 378, 255 

\bibitem -Koyama, K., Kinugasa, 
K., Matsuzaki, K., Nishiuchi, M., Sugizaki, M., Torii, K., Yamauchi, S., \& 
Aschenbach, B.\ 1997, PASJ, 49, L7 


\bibitem -Morrison, R.~\& 
McCammon, D.\ 1983, ApJ, 270, 119 

\bibitem 
-Reynolds, S.~P.~\& Keohane, J.~W.\ 1999, ApJ, 525, 368 

\bibitem -Slane, P., Plucinsky, P., Harrus, I.~M., Hughes, J.~P., Green, A.~J., \& Gaensler, B.~M.\ 1997, Bulletin of the American Astronomical Society, 29, 1368

\bibitem -Slane, P., Hughes, J.~P., 
Edgar, R.~J., Plucinsky, P.~P., Miyata, E., Tsunemi, H., \& Aschenbach, B.\ 
2001, ApJ, 548, 814 

\bibitem -Str{\" u}der, 
L.~et al.\ 2001, A\&A, 365, L18 

\bibitem -Sugizaki, M., Mitsuda, 
K., Kaneda, H., Matsuzaki, K., Yamauchi, S., \& Koyama, K.\ 2001, ApJS, 
134, 77 

\bibitem -Tanimori, T.~et al.\ 
1998, ApJL, 497, L25 

\bibitem -Turner, M.~J.~L.~et al.\ 
2001, A\&A, 365, L27 

\bibitem -Ueno, 
M., Bamba, A., Koyama, K., \& Ebisawa, K.\ 2003, ApJ, 588, 338 

\bibitem -Vink, J., 
Kaastra, J.~S., \& Bleeker, J.~A.~M.\ 1996, A\&A, 307, L41 

\bibitem 
-Vink, J., Kaastra, J.~S., Bleeker, J.~A.~M., \& Bloemen, H.\ 2000, Advances 
in Space Research, 25, 689 

\bibitem -Yamauchi, S.~et al.\ 
2002, 8th Asian-Pacific Regional Meeting, Volume II, 81 

\bibitem -Yamazaki, R., 
Yoshida, T., Terasawa, T., Bamba, A., \& Koyama, K.\ 2004, A\&A, 416, 595

\end{thebibliography}
\end{document}